# Chip-based Resonance Raman Spectroscopy Using Tantalum Pentoxide Waveguides

David A. Coucheron,[1] Dushan N. Wadduwage,[2,3] G. Senthil Murugan,[4] Peter T. C. So,[2,3] Balpreet S. Ahluwalia[1,*]

[1]Department of Physics and Technology, UiT - The Arctic University of Norway, 9037 Tromsø, Norway
[2]Laser Biomedical Research Center, Massachusetts Institute of Technology, Cambridge, Massachusetts 02139, USA
[3]Department of Biological Engineering, Massachusetts Institute of Technology, Cambridge, Massachusetts 02139, USA
[4]Optoelectronics Research Centre, University of Southampton, Southampton SO17 1BJ, United Kingdom

*Corresponding author: balpreet.singh.ahluwalia@uit.no

**Blood analysis is an important diagnostic tool, as it provides a wealth of information about the patient's health. Raman spectroscopy is a promising tool in blood analysis, but widespread clinical application is limited by its low signal strength, as well as complex and costly instrumentation. The growing field of waveguide-based Raman spectroscopy tries to solve these challenges by working towards fully integrated Raman sensors with increased interaction areas. In this work, we demonstrate on-chip resonance Raman measurements of hemoglobin, a crucial component of blood, at 532 nm excitation using a tantalum pentoxide ($Ta_2O_5$) waveguide platform. We have also characterized the background signal from $Ta_2O_5$ waveguide material when excited at 532 nm. In addition, we demonstrate spontaneous Raman measurements of iso-propanol and methanol using the same platform. Our results suggest that $Ta_2O_5$ is a promising waveguide platform for resonance Raman spectroscopy at 532 nm, and in particular for blood analysis.**

Blood is an essential body fluid that consists mainly of red blood cells, white blood cells and platelets in a liquid medium called plasma. The components of blood serve different purposes, ranging from oxygen transport to fighting disease and infection. Analyses of blood can thus give vital insight into a person's health and can identify diseases such as diabetes, malaria and sickle-cell disease [1]. Hemoglobin, the major component of red blood cells, has been of interest in the Raman spectroscopy community since the 1970s [2], as it dominates the Raman spectrum of blood. Traditional Raman spectroscopy has a very low scattering cross-section and is thus a weak effect. Hemoglobin, however, experiences resonance Raman at 532 nm excitation, enhancing the signal significantly [2]. The widespread application of Raman spectroscopy in routine clinical analysis is, however, limited by e.g. low throughput, complex and costly instrumentation [3].

On-chip Raman spectroscopy [4] using integrated optical waveguides is an attractive route for delivering compact, simple to operate, affordable and highly sensitive Raman spectrometers. The main reason is that the use of waveguides enables a simple way to increase the Raman scattering, as well as the possibility of a fully integrated sensor on chip, making instrumentation far simpler. Moreover, the mass production of integrated photonic chips can bring its cost significantly down.

In particular, Raman spectroscopy using waveguides made of high-refractive index contrast (HIC) materials is experiencing a renewed interest [5–8]. Recent achievements include measurements of biological monolayers [9] and trace gases [10]. When light is guided in an optical waveguide, an evanescent field is generated outside the waveguide core, which can be used to probe a sample close to the waveguide surface. The evanescent field intensity can be made sufficiently high for Raman spectroscopy by fabricating thin waveguides (100-150 nm) made of HIC materials. Raman scattering from the sample will either recouple into the waveguide or emit into the free space (as illustrated in Fig. 1a by the orange arrows). Detection can therefore be done either from the top of the waveguide or from the end facet of the waveguide-chip, as illustrated in Fig.1b. This technique is commonly called waveguide enhanced Raman spectroscopy (WERS) when detection is done from the end, as the waveguide structure gives rise to intrinsic enhancement effects for thin samples (e.g. monolayers). Waveguides made of HIC materials benefit from two enhancement effects when using edge detection for thin samples: 1) Radiative enhancement due to the high intensity of the evanescent field and 2) interaction area enhancement due to the increased interaction area [6]. The HIC material tightly confines the light within the waveguide core, enabling ultra-small footprints of the optical functions. High confinement allows sharp bends, significantly enhancing the interaction area by designing e.g. a spiral geometry in a small footprint [5]. If the propagation loss of the waveguide platform is low, arbitrarily large sensing area can be designed using waveguides made of HIC materials. The instrumentation can also be made much simpler, as a waveguide based set-up can have integrated lasers [11], detectors [12] and other components needed for Raman spectroscopy.

Most of the recent result on waveguide based Raman spectroscopy has been performed using silicon nitride ($Si_3N_4$) nanowires [5] and slot waveguides [13] with 785 nm excitation light. Spiral waveguides are highly sensitive and have been used to measure biological sub-monolayers [9]. Despite the success of $Si_3N_4$ spiral waveguides, getting sufficient Raman scattering without cooled detectors, which is relevant for a fully integrated solution, remains challenging. Several approaches have been taken to improve the signal further in combination with the waveguide platform, including surface enhanced Raman scattering [14] and Stimulated Raman Scattering [8].

Another option for increasing the Raman signal, is by moving to lower wavelengths, as Raman scattering scales with $\lambda^{-4}$. Other advantages of lower excitation wavelength are tighter confinement, and possible resonance effects for selective samples. Resonance Raman occurs for certain samples at low wavelengths and can enhance the signal by several orders of magnitude. Using 532 nm excitation is particularly interesting due to resonance Raman effects in several clinically relevant samples, such as hemoglobin [2] and in cancerous brain [15] and breast tissue [16].

The development of HIC waveguide platform for resonance Raman spectroscopy using 532 nm excitation is therefore beneficial. The $Si_3N_4$ waveguide platform can have a high background signal at shorter wavelength, which can be detrimental for Raman spectroscopy [17]. Evans et al. have recently demonstrated waveguide based Raman spectroscopy using a $TiO_2$ platform at 532 nm excitation [7]. Another potential platform for shorter wavelength excitation is $Ta_2O_5$. $Ta_2O_5$ is a promising waveguide material for Raman spectroscopy due to its high refractive index contrast with silica cladding, providing strong optical mode confinement and thereby enabling the development of compact photonic circuits with tight bends and spirals. $Ta_2O_5$ is CMOS compatible and mass-producible, and offers a wide transparency window with UV absorption band edge below 330nm and has been widely used in sensing applications [18]. Wang et al. have demonstrated both edge and top detection using $Ta_2O_5$ waveguides with 638 nm excitation [6], however, resonance Raman using 532 nm has yet to be investigated for biological application.

In this work, we present waveguide-based Raman spectroscopy using $Ta_2O_5$ waveguides using 532 nm excitation. We demonstrate spontaneous Raman spectroscopy of isopropanol and methanol. The background signal from the waveguide platform is investigated by changing the under cladding material between silicon dioxide ($SiO_2$) and magnesium fluoride ($MgF_2$). Finally, we measure resonance Raman spectrum of hemoglobin, suggesting the potential of $Ta_2O_5$ waveguide based Raman spectroscopy platform for biological applications.

A 532 nm laser (Verdi V10) was used for all experiments. The laser was coupled into the waveguides using a 0.5 N.A. aspheric lens on a three axis piezo stage. Collection was done from the top using a 60x/1.2 NA WI objective lens (Olympus), unless otherwise stated. All work was performed with top collection, as we aim to use spatial information in future work. The emission light was relayed through two 532 nm edge filters (Semrock RazorEdge) and coupled into a multimode fiber. A spectrograph (Andro Holospec-F/1.8-VIS) was used to analyze the signal coupled out of the fiber. A schematic of the set-up used is presented in Figure 1b. All sample measurements were done using $Ta_2O_5$ strip waveguides on oxidized silicon (Si) wafers. The optimization of the fabrication process is described in previous work [19].

Additionally, $Ta_2O_5$ waveguides were fabricated on $MgF_2$ substrates to study the background signal from the under-cladding material. Strip waveguides with different widths were defined in the $Ta_2O_5$ layer using standard photolithography and etched by 150 nm using argon ion beam milling. Finally, the waveguide samples were annealed at 600 °C for 3 hours under an oxygen atmosphere to relieve any stress built up in the films [20] and to replenish the oxygen, which is depleted during the sputtering process. 2.5 μm wide $Ta_2O_5$ waveguides on $SiO_2$ and on $MgF_2$ were used for the measurement. Prior to measurements, the waveguides were cleaned for 10 minutes in 1% Hellmanex in deionized (DI) water at 70°C. The waveguides were then rinsed in DI water, followed by isopropanol and DI water again. Pressurized air was used to dry the waveguides. A hollow rectangular PDMS chamber of a few micron thickness was attached to the waveguide when measuring liquid samples. The PDMS chamber defines the measurements area. The sample was then added in the chamber and a cover slip was used to seal it.

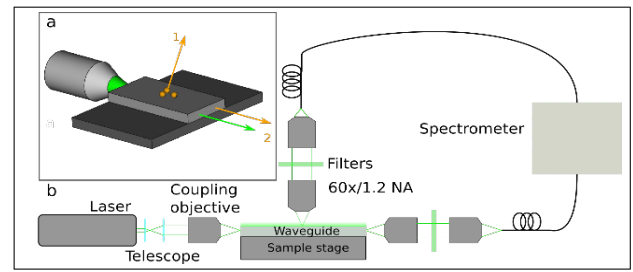

Figure 1: (a) Illustration of Raman scattering from a sample on a waveguide. The light guided in the waveguide generates an evanescent field stretching 100-200 nm outside the structure. Raman scattering is induced by the sample and the scattered photons can either be transmitted into free space (1) or recoupled back into the waveguide (2). The signal can thus be collected either from the top or the edge, where top detection gives some spatial information and edge detection gives a large area enhancement. The evanescent field will also penetrate into the substrate which can add to the background signal (b) Schematic of the system used in the experiments. Only results from top detection are used in the presented work, but edge detection was included in the schematic for completeness.

First, we measured the background spectrum of $Ta_2O_5$ waveguides. The measurements were done with a 70 μm wide $Ta_2O_5$ strip waveguide. A laser beam (38 mW power) was incident at the input facet of the waveguide. Measurements were taken using a 20 s exposure time. The resulting spectra were normalized with respect to the highest peak at approximately 680 cm-1 and is presented in Figure 2. $Ta_2O_5$ only exhibits a background signal from wavenumbers shorter than approximately 1000 cm-1. Four peaks can be discerned: a large peak at 680 cm−1; two smaller peaks at 488 cm−1 and 196 cm−1; and a shoulder at 899 cm−1. This agrees well with previous literature on amorphous $Ta_2O_5$ [20]. It is important to note that the $Ta_2O_5$ background is much weaker above approximately 1000 cm−1. Using suitable filters, sample spectra above 1000 cm−1 could be analyzed with only minimal background distortion.

We measured the background spectrum for $Ta_2O_5$ strip waveguides of different widths (2.5-70 μm) as shown in Fig. 2. The overall signal remained unchanged as the waveguide width was changed, but a new peak emerged at approximately 523 cm−1 for the narrower waveguides, which is typical Si-Si peak, originating from under-cladding. Fig. 2b presents the ratio of the peaks at 523 cm−1 to the local background level at 485 cm-1, normalized for 70 μm width. With decreasing width, the ratio increases rapidly below approximately 5 μm. For thinner waveguides, signals from the under-cladding becomes more prominent. For thin (150 nm) and narrow waveguides (<5 μm), the guided mode stretches further outside the core into the cladding, resulting in small contribution from the cladding. The results suggest that most of the overall spectral signal is from the $Ta_2O_5$ core and not the under-cladding for wide waveguides.

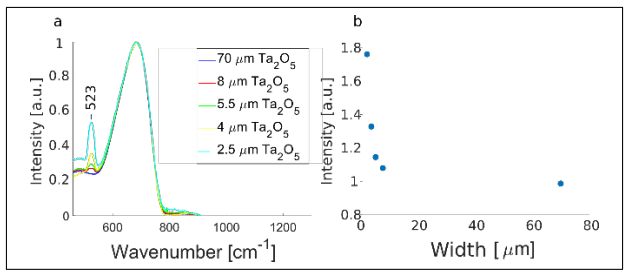

Fig. 2. (a) Normalized spectra measured as a function of waveguide width. (b) The ratio of the emerging peak at 523 cm$^{-1}$ to the largest peak at 680 cm$^{-1}$ as a function of waveguide width. The spectra are normalized to the widest waveguide.

To further investigate the signal originating from the under-cladding, we compared $Ta_2O_5$ waveguides fabricated with two different under-claddings, silicon dixode ($SiO_2$) and magnesium fluoride ($MgF_2$). Narrow $Ta_2O_5$ strip waveguide of width 2.5 μm were measured for two under-claddings, $SiO_2$ and $MgF_2$. A laser beam with 110 mW power was incident at the input facet of the waveguides and 10 s exposure time was used. The spectrum was averaged over 10 measurements to reduce the noise. The resulting spectra are presented in Fig. 3. The general shape of the spectrum remains unchanged, suggesting that the waveguide material contributes to most of the background signal. There are, however, a few notable changes. The sharp peak at 523 cm$^{-1}$ from the $Ta_2O_5$ on $SiO_2$ is replaced with two new peaks at 410 cm$^{-1}$ and 295 cm$^{-1}$ for the $Ta_2O_5$ on $MgF_2$. The peaks at 410 cm$^{-1}$ and 295 cm$^{-1}$ are confirmed to be Raman peaks from $MgF_2$ [21]. Reducing the background further is thus challenging, since the background signal is dominated by the waveguide material and cannot easily be changed. It might, however, be possible to reduce the background by changing the fabrication process and a more detailed study into the origin of the signal in the $Ta_2O_5$ structure is needed.

Spontaneous Raman measurements of methanol and isopropanol with 532 nm excitation were performed using 150 nm thick $Ta_2O_5$ strip waveguides. The waveguides were 70 μm wide to increase the sensing area. A laser beam of 395 mW power was launched from the coupling objective. Both isopropanol and methanol were measured using a PDMS chamber to hold the sample, sealed with a cover glass. An exposure time of 20 s was used for both the samples. The presented spectrum is the average over three measurements. Spectra for methanol and isopropanol using waveguide excitation are presented in Fig. 4a. The waveguide background has been subtracted and all spectra have been normalized to the 822 cm$^{-1}$ peak for isopropanol and the 890 cm-1 peak for methanol. The peaks agree well with literature [22] and the two chemicals are easily distinguishable by e.g. the symmetric C-O-O vibration at 822 and 890 cm$^{-1}$ for isopropanol and methanol, respectively.

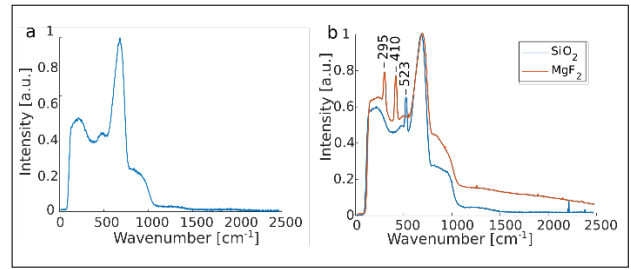

Figure 3: (a) Background spectrum from a 70 μm wide $Ta_2O_5$ strip waveguide on a $SiO_2$ undercladding. (b)Background spectrum from 2.5 μm wide $Ta_2O_5$ waveguides on a $SiO_2$ and $MgF_2$ undercladding. The overall shape of the spectrum remains unchanged for the different undercladdings, but some characteristic peaks change.

A motivation for using 532 nm excitation is to measure resonance Raman signals from hemoglobin, as it is important in lab-on-a-chip blood analysis for point of care diagnosis. Resonance Raman can distinguish different hemoglobin derivatives accurately [2]. By using waveguide enhanced Raman, instrumentation can be simplified and costs reduced. For hemoglobin resonance Raman we used human hemoglobin (lyophilized powder from Sigma Aldrich) with 15.5 mM concentration in distilled water, which is similar to the hemoglobin concentration in red blood cells [23]. A drop was placed on the waveguide surface and allowed to dry. Collection was done with a 40x/0.75 NA air objective lens for this experiment, with 2 s exposure time.

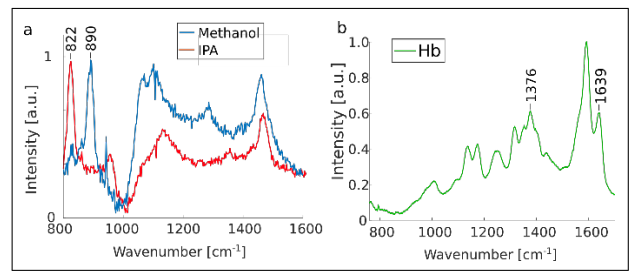

Figure 4: (a) Spontaneous Raman spectra of methanol and isopropanol (IPA). (b) Resonance Raman spectrum of dried hemoglobin (Hb).

Fig. 4b shows the resonant Raman spectrum of dried hemoglobin. The Raman spectrum of hemoglobin is complex and exhibits a numerous Raman peaks [1], but the presence of peaks at 1639 cm$^{-1}$ and 1376 cm$^{-1}$ demonstrates that the sample is oxygenated. The Raman peaks are mostly above 1000 cm$^{-1}$, therefore, they are not overlapping with the waveguide background. This makes $Ta_2O_5$ ideally suited for resonance Raman spectroscopy for hemoglobin and blood analysis for point of care diagnostics. Moving to lower wavelengths (e.g. 532 nm) increases the Raman signal significantly and several interesting samples exhibit resonance Raman at visible wavelengths. Clinically relevant samples, such as breast tissues [15] and brain tissues [14] experience resonance Raman effects at 532 nm excitation. Here, we have employed only straight strip waveguides and top detection. The sensitivity can be further increased by changing to a spiral loop geometry and by doing edge detection [6], which will be explored in future work.

Photonic integrated circuits provide a promising approach to Raman spectroscopy. $Ta_2O_5$ waveguides with its low background signal and high refractive index contrast has been demonstrated as a suitable platform for resonance Raman spectroscopy at 532 nm. By investigating waveguides of different under-claddings it was found that most of the background comes from the $Ta_2O_5$ core and only a small background from the surrounding media is visible for narrow waveguides. We have tested and validated the waveguide platform by measuring spontaneous Raman spectra of isopropanol and methanol. We have performed resonance Raman measurements of dried hemoglobin at concentration usually found in the red blood cells [21] using the $Ta_2O_5$ waveguides. Waveguide based Raman spectroscopy has the potential to bring Raman spectroscopy into the clinic by offering a cheap and small instrument, where a disposable photonic chip can be used both to hold and to illuminate the sample. The resonance Raman spectroscopy can also be combined with other complementary optical functions developed using $Ta_2O_5$ material, such as optical nanoscopy [24] and optical trapping [25].

**Funding.** European Research Council (project number 336716) National Science Foundation (NSF) (1263236, 0968895, 1102301); The 863 Program (2013AA014402)

**Acknowledgment**. B.S.A acknowledges the funding from the European Research Council, (project number 336716). B.S.A and D.A.C acknowledges UiT, The Arctic University of Norway Tematiske Satsinger funding program.